\documentclass[floatfix,aps,prl,twocolumn,showpacs,superscriptaddress,groupedaddress]{revtex4-1}  
\usepackage{graphicx}  
\usepackage{dcolumn}   
\usepackage{bm}        
\usepackage{amssymb}   
\usepackage{amsmath}
\usepackage{appendix}
\usepackage{color}
\usepackage[normalem]{ulem}  
\usepackage[abbreviations]{foreign} 
\usepackage{xr-hyper}
\usepackage{hyperref} 
\usepackage[all]{hypcap} 
\makeatletter
\newcommand*{\addFileDependency}[1]{
  \typeout{(#1)}
  \@addtofilelist{#1}
  \IfFileExists{#1}{}{\typeout{No file #1.}}
}
\makeatother

\newcommand*{\myexternaldocument}[2][]{%
    \externaldocument[#1]{#2}%
    \addFileDependency{#2.tex}%
    \addFileDependency{#2.aux}%
}

\myexternaldocument[SI-]{Supplementary}

\begin{document}

\title{Assessing the polymer coil-globule state from 
the very first spectral modes}

\author{Timothy F\"oldes$^1$}
\email{timothy.foldes@sorbonne-universite.fr}
\author{Antony Lesage$^2$}
\author{Maria Barbi$^1$}
\email{maria.barbi@sorbonne-universite.fr}
\affiliation{$^1$ Sorbonne Universit\'e, CNRS, LPTMC, F-75005 Paris, France}
\affiliation{$^2$ Sorbonne Universit\'e, CNRS, PHENIX, F-75005 Paris, France.}

\date{\today}

\begin{abstract}

The determination of the coil-globule transition of a polymer is generally based on the reconstruction of scaling laws, implying the need for samples from a rather wide range of different polymer lengths $N$. The spectral point of view developed in this work allows for a very parsimonious description of all the aspects of the finite-size coil-globule transition on the basis of the first two Rouse (cosine) modes only, shedding new light on polymer theory and reintroducing well-established spectral methods that have been surprisingly neglected in this field so far. Capturing the relevant configuration path features, the proposed approach enables to determine the state of a polymer without the need of any information about the polymer length or interaction strength. Importantly, we propose an experimental implementation of our analysis that can be easily performed with modern fluorescent imaging techniques, and would allow differentiation of coil or globule conformations by simply recording the positions of at least three discernible loci on the polymer.
Furthermore, the framework we put forward applies to virtually any polymer model, fractal globule, loop extrusion...

\end{abstract}

\maketitle

The coil-to-globule phase transition, which describes the abrupt compaction of a polymer due to environmental changes, characterizes the physics of polymers in solution, with direct consequences for macroscopic solution properties such as viscoelasticity or transport features and, moreover, successfully models the heterogeneity of chromatin density, which most likely plays a crucial functional role in the regulation of gene expression.

It is usually studied by means of the scaling properties of the radius of gyration as a function of the number of monomers, $N$.
However, this approach requires comparing polymers of different lengths, which is not always possible. Even the determination of the number of monomers $N$ is not always easy, as it depends on the Kuhn length of the polymer, hence its bending persistence, which can be difficult to determine in the case of real polymers, and a fortiori for complex biological polymers. Moreover, important deviations from the theoretical scaling are induced by the finite size of the polymers~\cite{Lesage2019}.

From a different perspective, the dynamics of the polymer can be studied.  Typically, the dynamics of single monomers can be compared to theoretical predictions such as those of the Rouse or Zimm models~\cite{DoiEdwards, Socol2019}. 
Now, interestingly, the Rouse model uses a Fourier-like decomposition of the polymer conformation. Noting that this decomposition applies to any given conformation, we used it to determine the average spectral content of the polymer to inquire its equilibrium properties and study the coil-globule transition, rather than its dynamics. 
This paper aims to investigate this issue and to see how a spectral representation can facilitate the determination of the polymer state (\ie{} its macrostate).
We developed a method to discriminate the folding state of a polymer of a given size $N$, without the need to study the scaling law or to have access to the energy parameter.
Our method exploits the information about the spatial path of the polymer and only relies on \emph{low} spectral modes, \ie{} long-distance features. 
The introduction of a new $N$-independent order parameter provides us with a parsimonious and robust measurement of the state of the polymer on the coil to globule scale. Finally, based on this analysis, we propose an innovative {experimental} approach to determine the equilibrium state of a polymer from fluorescence imaging, provided that a minimum of three loci, equally spaced along the chain, can be discernibly labeled.

\begin{figure*}[t]
    \centering
    \includegraphics [width = 1\textwidth]{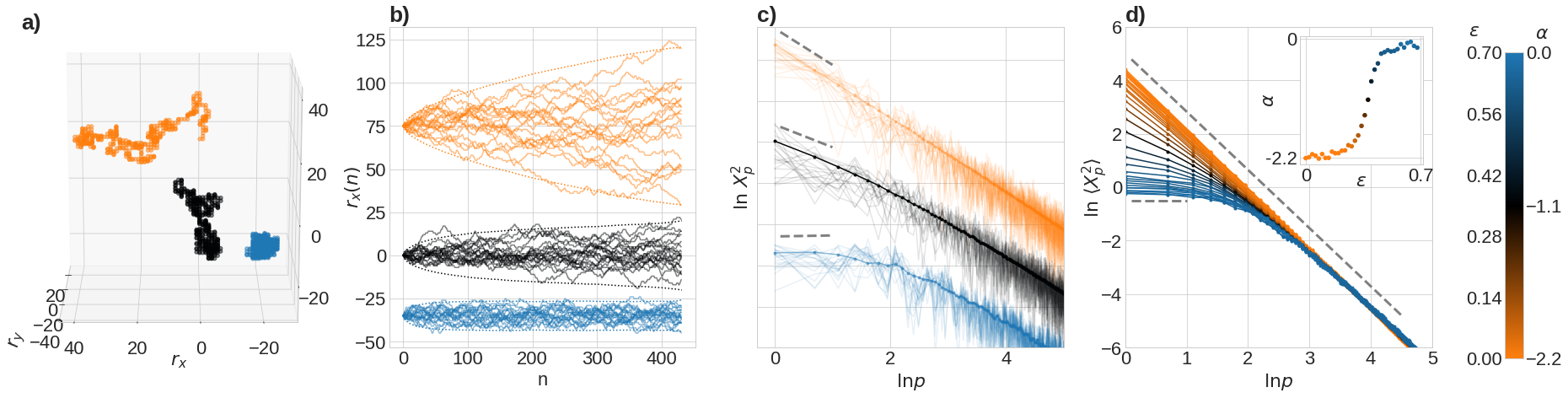}
    \caption{
    \textbf{(a)} On lattice MC simulation snapshots for $N= 431$ and $\varepsilon = 0$ (orange), $\varepsilon = 0.36$ (black) and $\varepsilon = 0.8$ (blue).
    \textbf{(b)} Projections $r_x(n)$ along the spatial $x$-axis of a set of 20 configurations, $n$ being the monomer index. Colors refer to the same parameters as in (a). Dotted lines correspond to the $\pm  2.5 \ \sigma_x (n)$ envelopes. 
    \textbf{(c)} Logarithm of the square Discret Cosine Transform of the 20 trajectories of (b), $\ln |X_p|^2$, as functions of the logarithm of the mode number $\ln p$ (thin lines) and their average calculated over the ensemble of 16384 simulated configuration and over the three spatial directions (thick lines and dots). Standard deviations are smaller than the symbol size. Groups of curves are vertically shifted for aim of clarity.  Dashed lines correspond to slopes $  -2.2 \equiv -(1+2\nu)$, $-1.3$ and $0$ (from top to bottom).
    \textbf{(d)} Averaged Power Spectral Density (PSD) obtained by applying the same procedure as in (c) to equivalent simulations with $\varepsilon$ going from $0$ to $0.6$ (see colorbar).  Dashed lines correspond to slopes $-(1+2\nu)$ (top) and $0$ (bottom). \textbf{Inset:}  The slope $\alpha_N(\varepsilon)$ of the log-log representation of the PSD, estimated from the first 2 modes, as a function of $\varepsilon$.  
    }
    \label{fig:method}
\end{figure*}

The Rouse model describes an ideal chain (no excluded volume nor attractive interaction) for which the radius of gyration scales as the variance of a random walk, $\langle R^2\rangle  \sim N$. The chain dynamics is described by applying the overdamped Langevin equation to a bead-spring polymer of $N+1$ monomers, immersed in a fluid of viscosity $\gamma$ at temperature $T$~\cite{Rouse1953}. The application of the Langevin equation yields a system of $N+1$ three-dimensional coupled stochastic equations~\cite{DoiEdwards}. 
To solve the model, this system is decoupled by introducing the new set of variables $\{\vec{X}_p(t),\,p = 0\dots N\}$ 
\begin{equation}
\label{eq:Xp}
   {\vec{X}_p(t) = \frac{1}{N+1}\sum_{n=0}^N \vec{r}_n(t)\cos\left(\frac{p \pi}{N+1}\left(n + \frac{1}{2}\right)\right)}
    \end{equation}
called \textit{Rouse modes} of the polymer.
The $p=0$ mode corresponds to the position of the polymer center of mass. In the following, we will always focus on modes $p=1\dots N-1$ only. Usually, one uses the auto-correlation function of $\vec{X}_p(t)$ to derive single monomer dynamical scaling laws, overlooking the potentialities of the Rouse decomposition in itself.
Indeed, in the language of signal processing, each component of \eqref{eq:Xp} is a Discrete Cosine Transform (DCT), closely related to the Discrete Fourier Transform (DFT), in that it corresponds to the DFT of a symmetrized signal of double length~\cite{Makhoul1980}.
Hence, the average over a set of equilibrium configurations
of the square mode amplitudes $\langle X^2_{p}\rangle$ is none other than the canonical estimator of the \emph{Power Spectral Density} (PSD), characterizing a polymer macrostate in terms of the relative weight of the different spectral modes.
For the Cartesian coordinate $x$ (equivalently $y$ or $z$), it is well known that~\cite{Rouse1953,DoiEdwards}
\begin{equation}
     \langle X^2_{p,x}\rangle = \frac{ b^2 }{24\, (N+1) \sin^2 \left(\frac{p \pi}{2(N+1)}\right)}  \quad \underset{p\ll N}{\sim}\quad \frac{1}{p^2}, \label{modesamplitude}
\end{equation}
yielding an explicit expression for the mean square amplitude of the Rouse modes of an ideal polymer.

The connection between Rouse modes and PSD seems, surprisingly, to have never been noticed before. However, the asymptotic dependence~\eqref{modesamplitude} has an immediate interpretation in terms of the equivalence to Brownian motion trajectories~\cite{Grosberg1994}, for which the PSD has a scaling exponent $-2$. Interestingly, exactly the same expression~\eqref{modesamplitude} is obtained in a different setting~\cite{Krapf2018}.

Now, a real polymer is characterized by interactions between monomers: repulsive excluded volume and (solvent dependent) attractive interactions.
The case where only excluded volume is present defines the (non-interacting) \emph{self-avoiding polymer} model, \ie{} the prototypical coil polymer. Due to steric hindrance, the polymer occupies more space resulting in swollen conformations analogous to self-avoiding walks (SAW). The polymer's radius of gyration then scales as $\langle R^2\rangle  \sim N^{2\nu}$, where $\nu \approx 0.588$ is the Flory exponent~\cite{Grosberg1994}.
Including steric interactions adds additional non-linear coupling terms to the Rouse equation system, and, consequently, there is no explicit solution for the PSD~\cite{Panja2009}. Yet simple arguments provide insight into the general trend of the spectrum. A usual (yet unproven) assumption about SAWs is their fractal nature, meaning that their statistical properties are scale invariant~\cite{Grosberg1994, Slade2019, FractalsInPhysics}.  Hence, we expect a power law PSD, intuitively related to the presence of long-range correlations induced by steric interactions.
The Hurst exponent $H$ provides an estimation of the long-range memory of a signal. In particular, for a centered fractal signal of Hurst index $H$ one has $\langle r_n^2\rangle \propto n^{2H}$ and a PSD scaling as $p^{-(1+2H)}$. By guessing a long-range correlated process for the SAW, the comparison with the Flory scaling $\langle R^2\rangle  \sim N^{2\nu}$ would lead to the ansatz $H = \nu$ which appoints the PSD
\begin{equation}
    \langle \vec{X}_p^2 \rangle 
    \propto p^{-(1+2\nu)}
    \label{eq:PSDSAW}
\end{equation}
where $-(1+2\nu) \approx -2.2$. 
\citeauthor{Panja2009} made an equivalent ansatz for long wavelengths Rouse modes of the self-avoiding polymer, and corroborate it by on-lattice simulations~\cite{Panja2009}.

Depending on the physico-chemical properties of the polymer and the solvent, an effective attraction $J$ between monomers may also exist. The state of the polymer thus depends on the relative strength of inter-monomer forces and thermal energy, $\varepsilon = {J}/{k_B T}$. This case is often modeled by the Interacting Self-Avoiding Walk (ISAW), an on-lattice polymer with each nearest neighbor monomer pair contributing $-\varepsilon$ to the energy (the SAW model is retrieved for $\varepsilon=0$).
In the thermodynamic limit $N\to \infty$, above a critical value $\varepsilon_\theta$ ($\theta$-point), the strong attraction induces a second order phase transition to curled up conformations of uniform density called globules, whose typical scaling is $\langle R^2\rangle \propto N^{2/3}$~\cite{Grosberg1994}. However, {finite-size} polymers undergo a smooth coil-globule transition at a {$N$-dependent} critical energy~\cite{DeGennes1979, Grassberger1995, Vogel2007, Care2014}.

The spatial configuration of the chain can be described by noting that a collapsed polymer is equivalent to an ideal polymer compressed within a spherical volume of radius $R$~\cite{Grosberg1994}.
Over small length scales, the polymer behaves like a free chain, until it reaches the volume boundary, where it is reflected and starts an independent random walk. 
The chain can thus be described as a series of independent subchains. The positions of two monomers in different sub-chains are uncorrelated~\cite{Grosberg1994}, resulting in the sequence of positions of monomers sufficiently spaced along the chain being a white noise signal.
From the spectral point of view, this implies a constant spectrum for small values of $p$. We thus predict that the PSD of globules verifies the property
\begin{equation}
    \langle \vec{X}_p^2 \rangle 
    \propto p^{0} \quad \quad \mbox{for small }p.
    \label{eq:flat}
\end{equation}

\begin{figure}[t]
    \centering
    \includegraphics [width = 1\columnwidth]{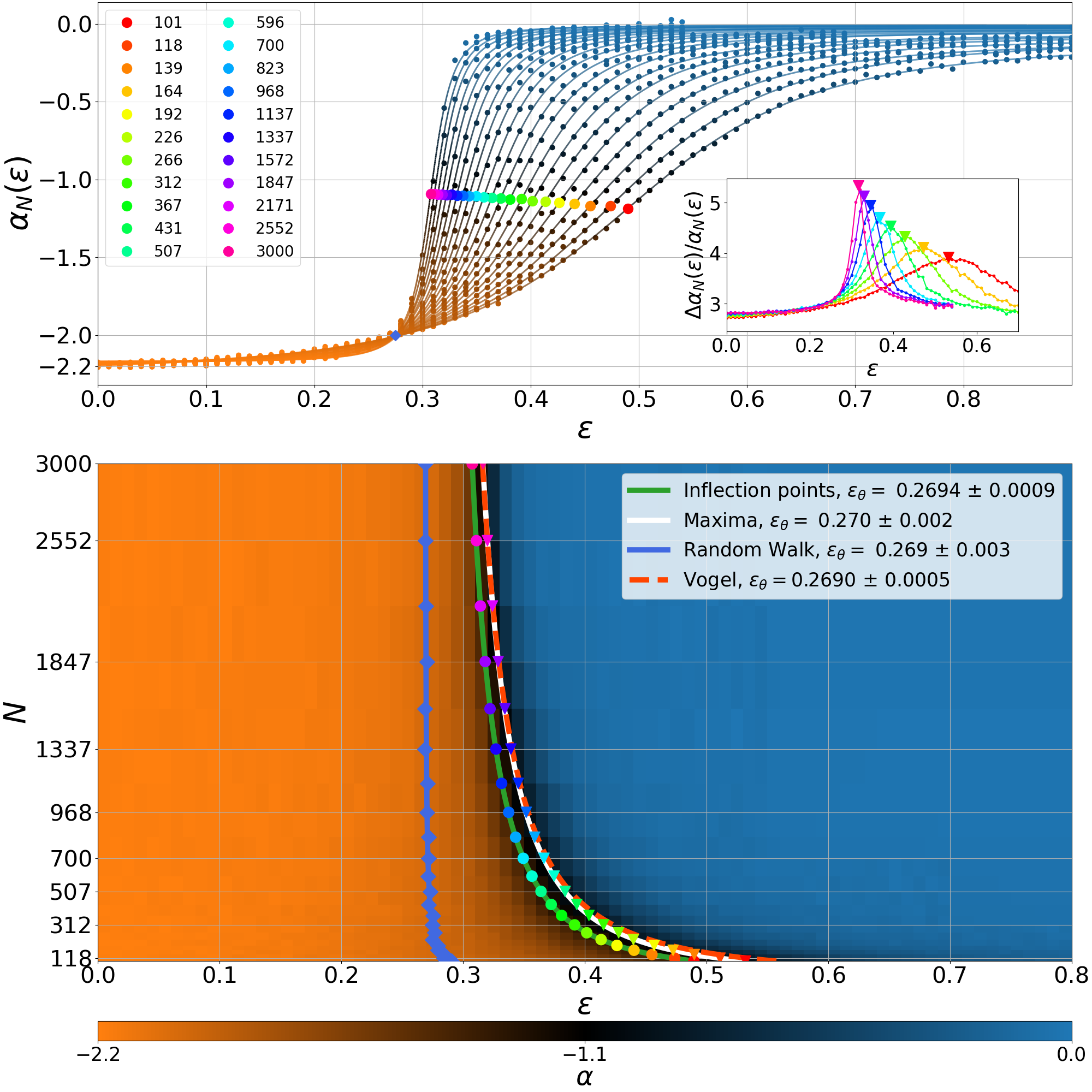}
    \caption{ \textbf{Top panel:}
    Log-log plot slope $\alpha_N(\varepsilon)$ estimated on the first 2 modes of  Power Spectral density (dots) for all polymers sizes $N$ from 101 to 3000 (right to left). For each $N$, $\alpha_N(\varepsilon)$ is fitted by a sigmoid ${S(x)=D + A(x - B)/\sqrt{1 + C(x - B)^2}}$ (lines). 
    Colored circles are inflection points, defining $\varepsilon^I(N)$. Blue diamonds indicate the $\alpha_N(\varepsilon) = -2$ condition, defining $\varepsilon_\theta^{RW}(N)$.  \textbf{Inset:} $\alpha_N(\varepsilon)$ relative standard deviation, ${\Delta \alpha_N(\varepsilon)}/{\alpha_N(\varepsilon)} = {\sigma_{X_1^2}}/{\langle X_1^2 \rangle} + {\sigma_{X_2^2}}/{\langle X_2^2 \rangle}$.
     Triangles indicate the position of maxima $\varepsilon^F(N)$, determined by fitting the top of the pics by a Gaussian function. 
     \textbf{Bottom panel:} Corresponding phase diagram in the $(\varepsilon, N)$ phase space. The orange to blue colorscale correspond to values of $\alpha_N(\varepsilon)$ from -2.2 to 0 (see colorbar). Inflection points (colored circles) and fluctuation maxima (colored triangles) are shown and fitted by curves of expression $\varepsilon_\theta(N) = {a_1}{{N^{-1/2}}} + {a_2}{(2N)^{-1 }} + \varepsilon_\theta$ (green and white lines, respectively). The red dashed line is the critical curve of same expression, with the parameters  obtained by \citeauthor{Vogel2007} in Ref.~\cite{Vogel2007}. Fitting parameters are given in Table S1. 
     \textit{Both panels}:  Samples from the sectors $(N, \varepsilon) \in [700;3000] \times [0.55;1.0]$ and $ [300;700] \times [0.7;1.0]$ didn't meet statistical relevance and were omitted from the results. }
    \label{fig:phasediagram}
\end{figure}

To test our hypotheses, we simulated single ISAW on the cubic lattice~\cite{Tesi1996} and sampled conformations thanks to the Metropolis algorithm with reptation moves~\cite{Wall1975}. 
We performed simulations for $N$ ranging from 100 to 3000, and $\varepsilon$ in the range  0.0--0.99. For each  $(N, \varepsilon)$, 32000 statistically independent conformations are recorded. \autoref{fig:method}-a shows typical conformations for the case $N=431$ and three different $\varepsilon$.
The three spatial components are DCT transformed and their square amplitudes are summed to obtain their DCT spectra, as shown in \autoref{fig:method}-c. 
For any given $(N, \varepsilon)$, we average over the set of configurations to obtain the corresponding PSD (thick lines in \autoref{fig:method}-c).
\autoref{fig:method}-d shows the resulting PSDs for different values of $\varepsilon$ and for the given $N$.

The results confirm both our predictions. First, for small values of the interaction parameter $\varepsilon$, the PSDs display the expected scaling of Eq.~\eqref{eq:PSDSAW}, corresponding to a slope of $-(1+ 2 \nu) \approx -2.2$ in the log-log representation. 
Second, in the large $\varepsilon$ region, the spatial trajectories clearly show the effect of confinement into the globule volume (\autoref{fig:method}-b). Correspondingly, the low $p$ modes are strongly attenuated, leading to a rather flat spectrum, in agreement with our prediction~\eqref{eq:flat}. 

We aimed to compare the scaling of $\varepsilon_\theta(N)$ to previous results, typically defined either by the vanishing of the second virial coefficient~\cite{Grassberger1995}, by the condition of divergent specific heat~\cite{Vogel2007}, or by a specific scaling of $\langle R^2\rangle$~\cite{Rampf2006}.
In Ref.~\cite{Vogel2007}, \citeauthor{Vogel2007} show by on-lattice simulations that the boundary between the two phases is best fitted by a function of the form  $f(N) = {a_1}{{N^{-1/2}}} + {a_2}{(2N)^{-1}} + \varepsilon_\theta$.
We here propose two different definitions for the phase transition critical line. The first one is simply given by the inflection points of the sigmoid functions. The corresponding $\varepsilon^I(N)$ points are very well-fitted by $f(N)$ (\autoref{fig:phasediagram}, circles, and Table S1). 
A second, more physical definition of a transition line is
given by the $\varepsilon^F(N)$ that maximizes the fluctuations of our order parameter $\alpha_N(\varepsilon)$. Indeed, the relative standard deviation ${\Delta \alpha_N(\varepsilon)}/{\alpha_N(\varepsilon)}$ displays a sharp peak at the transition (meaning that the conformation of the polymer itself largely fluctuates) as shown by the inset of \autoref{fig:phasediagram}. The corresponding $\varepsilon^F(N)$ maxima are again best fitted by $f(N)$ (\autoref{fig:phasediagram}, triangles and Table S1).

Interestingly, both $\varepsilon^I(N)$ and $\varepsilon^F(N)$ correspond to values of $\alpha$ close to $\sim -1.1$ (black region) and are therefore far from $\alpha = -2$, associated with RW-like configurations usually identified with the $\theta$-point conditions. Our results show, instead, that these RW-like conditions do not appear at the crossover, but are rather located in the coil region of the phase diagram. More precisely, the condition $\alpha_N (\varepsilon_\theta) = -2$ is met at about $\varepsilon = 0.27$ (blue diamonds in \autoref{fig:phasediagram}) and indeed corresponds to a PSD decreasing as $p^{-2}$ over a wide $p$-region
(Figure S1).
The corresponding $\varepsilon^{RW}_\theta(N)$ curve is also fitted by a $f(N)$ function (although with negligible $a_1$, see
Table S1;
blue line in \autoref{fig:phasediagram}).
As expected, however, the asymptotic $\varepsilon_\theta$ is virtually the same for the three critical curves $\varepsilon^{I}_\theta(N)$, $\varepsilon^{F}_\theta(N)$ and $\varepsilon^{RW}_\theta(N)$ and converge to the well established thermodynamic limit value of $\varepsilon_\theta = 0.2690$, in agreement with previous numerical estimates~\cite{Grassberger1995,Vogel2007}.

The existence of two distinct critical lines was already pointed out by \citeauthor{desCloizeaux1991}~\cite{desCloizeaux1991}, but not clearly illustrated by numerical experiments, to the best of our knowledge. As a corollary message of our work, we confirm therefore that the conformation of a polymer of finite-size at the coil-globule transition is not reducible to a pure random walk. The very accurate phase diagrams obtained by spectral analysis validates it as a new, efficient and parsimonious method in polymer studies, neglected until now despite its very classical basis.

\begin{figure}[t]
    \centering
    \includegraphics [width = 1\columnwidth]{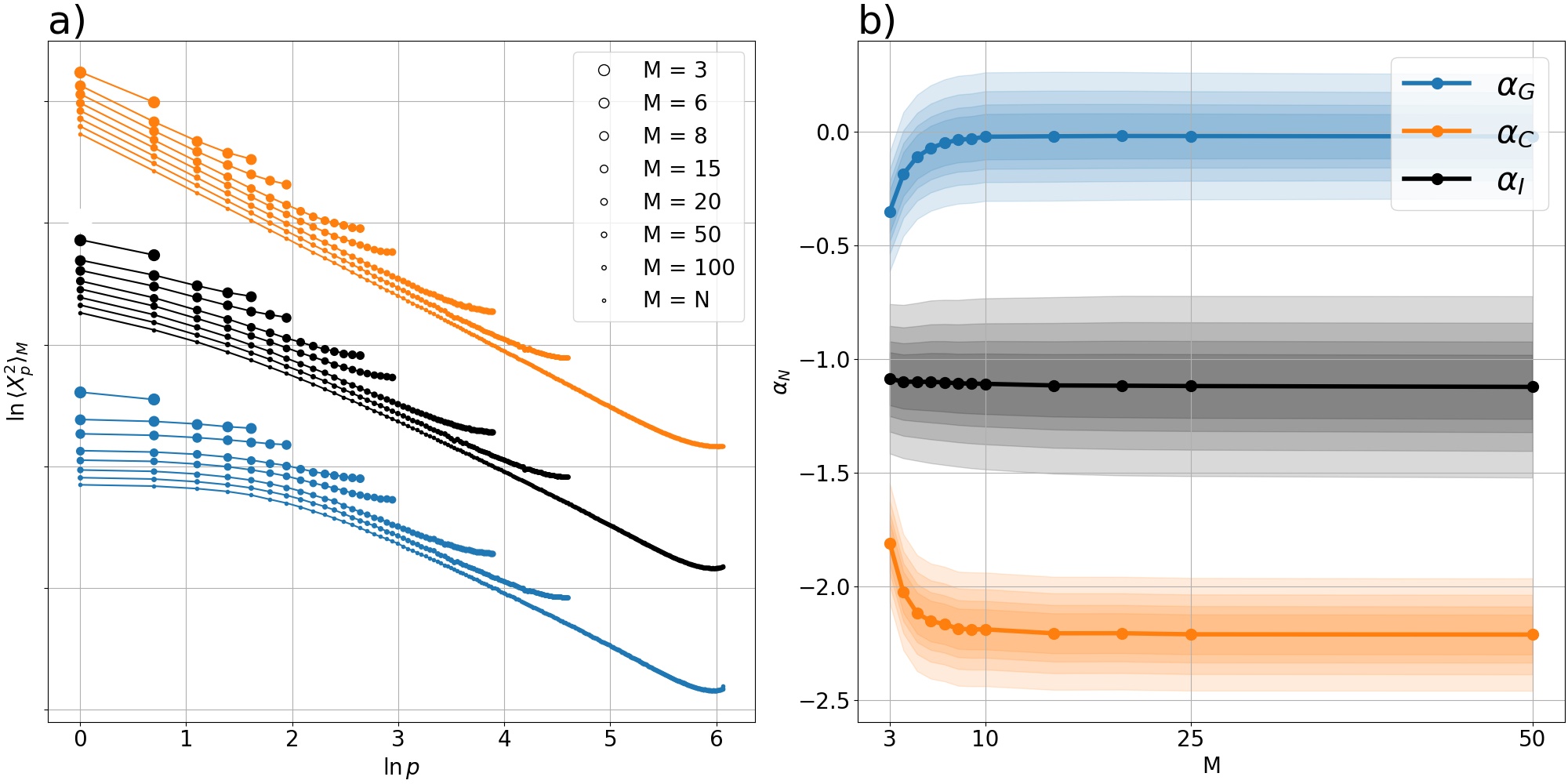}
    \caption{
    \textbf{(a)} Power Spectral Densities obtained by successive decimation of a $N= 431$ chain for $\varepsilon = 0.0$ (orange, coils), $\varepsilon = 0.32$ (black) and $\varepsilon = 0.46$ (blue, globules). Decimation goes from $M=N$ to $M=3$ (longer to shorter curves).
    \textbf{(b)} The asymptotic and critical values of $\alpha_{N,M}$ obtained from the sigmoid fit of the decimated spectra, as a function of $M$ (dots and lines): $\alpha_C(M) = \alpha_{N,M}(0)$ for coils, in orange; $\alpha_I(M) = \alpha_{N,M}(\varepsilon^I)$ for inflection points, in black; $\alpha_G(M) = \alpha_{N,M}(\varepsilon \to \infty)$ for globules, in blue. These values are averaged over $N$.
    Shaded regions correspond to $2\sigma_\alpha$ confidence intervals for different sample size from 256 (larger intervals) to 2048, calculated by propagating the statistical variance through Eq.~\eqref{eq:alpha}.
    We used a dataset of 16384 configurations for each $(N,\varepsilon, M)$ condition.
    }
    \label{fig:decimation}
\end{figure}

With our general hypotheses confirmed, we address the important issue of characterizing the coil-globule transition.
Inspired by our previous observations on the low $p$ limit,  we define as a new {order parameter} the log-log slope at $p=1$
\begin{equation}
    \alpha_N (\varepsilon) =\frac{\ln \langle X_2^2 \rangle - \ln \langle X_1^2 \rangle}{\ln 2 - \ln 1} = \frac{1}{\ln 2} \ \ln  \frac{\langle X_2^2 \rangle}{\langle X_1^2 \rangle}. 
    \label{eq:alpha}
\end{equation}
For the case of \autoref{fig:method}-d, $\alpha_N (\varepsilon)$ clearly performs a {continuous} transition from the typical SAW exponent $\alpha = -(1+2\nu)$ to $0$ (inset). The width of the transition region depends, however, on $N$. The $\alpha_N(\varepsilon)$ curves, for all $\varepsilon$ and $N$, have been fitted by a 4-parameter sigmoid function (\autoref{fig:phasediagram}, top panel), and all $\alpha_N(\varepsilon)$ indeed interpolate between the two limiting values $\alpha = -(1+ 2 \nu)$ and  $\alpha=0$. The fact that these limiting values are independent of $N$ and $\varepsilon$ makes of $\alpha_N(\varepsilon)$ an excellent order parameter for assessing the polymer state, as it can be applied to each {single} $(\varepsilon, N)$ polymer configuration set and {without the need of any information about these two parameters}. The bottom panel of \autoref{fig:phasediagram} shows a color plot of $\alpha$ as a function of $N$ and $\varepsilon$ where colors span from orange, for $\alpha = -2.2$, to blue for $\alpha =0$. The typical finite-size crossover region (in black) 
is clearly visible, allowing the definition of a critical transition line $\varepsilon(N)$. 

As a last, but most important point, the fact that our order parameter can be calculated on the basis of the first two spectral modes opens the possibility of {experimentally} assessing the state of a polymer from a very reduced and accessible information.
The argument relies on a well-known property of the PSD: In order to get access to the first $M$ modes, it is indeed in principle sufficient to record the positions of $M$ {distinguishable} monomers, {equally spaced} along the polymer and {covering the whole chain}, meaning that the determination of $\alpha$ would in theory require the knowledge of only 3 positions along the full polymer.

We tested the efficiency of our approach on \textit{decimated} polymer chains, \ie{} reduced signals where the positions $\vec r_n$  of only $M$ equally spaced monomers are retained. We decimated down to $M=3$, an extreme condition of particular interest from the experimental point of view   since, in this case,  time tracking can also be achieved without much difficulty~\cite{Chen2018,Sato2020,Eykelenboom2020} so that time averages can be used instead of ensemble averages.
\autoref{fig:decimation}-a shows examples of PSD obtained for different choices of $M$. Extreme decimation ($ M < 10 $) alters the first modes. As a consequence, the asymptotic values of the sigmoidal $\alpha_{N,M}(\varepsilon)$ corresponding to "pure" coils ($\alpha_C$) and globules ($\alpha_G$) vary, so that it is necessary to provide reference values for these limit slopes as a function of $M$.  We give these values in \autoref{fig:decimation}-b and in the Table S2. Noteworthy, the asymptotic values for coils and globules remain well apart down to $M=3$ even for relatively low statistical sampling meaning that the arrangement of monomers on large scales is detectable whatever the sampling.

Once these $M$-dependent limit values obtained, the order parameter $\alpha_{N,M}(\varepsilon)$ can be normalized as 
$    \tilde{\alpha}_N(\varepsilon) = 
    \big[{2 \alpha_{N,M}(\varepsilon)-\big( \alpha_G(M) + \alpha_C(M) \big)}\big]/
    {\big( \alpha_G(M) - \alpha_C(M) \big)}$
so to span from -1 to 1. In this way, equivalent sigmoids $\tilde{\alpha}_N(\varepsilon)$ are independent of $M$ Figure S2.
For sufficiently large samples, even an extreme decimation with $M=3$ allows a very accurate reconstruction of the $(N,\varepsilon)$ phase diagram (Figure S3). The critical line $\varepsilon^I(N)$ matches that of the complete chain (white line in Figure S3 and Table S3). The robustness of the proposed approach against variations in the size of the statistical sample is explored in Figure S4 showing that the universal character of the $\tilde \alpha_N$ parameter potentially outperforms any other method aiming at determining in which phase a polymer in given conditions falls.

These results have relevant implications in the context of the study of chromatin.
The characterization of the multiscale folding state of chromosomes is a crucial, issue since it likely determines its the biological activity. This is made evident from the strong progression of experimental methods to characterize this folding and their impact on the understanding of genetic regulation mechanisms~\cite{Jerkovic2021}. Very recently, microscopy-based techniques allow for the first time the visualization of this polymer spatial trajectory by sequential labeling and imaging of multiple loci in a collection of \textit{fixed} nuclei~\cite{CardozoGizzi2019,Mateo2019,Nguyen2020,Su2020}. This results in configuration sets, sampled with a resolution ranging from 3 to several hundreds of points, which are for the moment difficult to analyze, the most frequent approaches being the reconstruction of 
parameters that were already obtainable with previous techniques. 
There is therefore a clear need for efficient methods to process this new data to the fullest and without loss of information.
Our method of spectral analysis precisely fills this gap with a physically proven approach.
Finally, we underline the very broad scope of this spectral approach, which extends far beyond the pure coil and globule. One can easily predict a specific spectral signature for fractal globules, loops, stretched polymers or a variety of multifractal configurations, potentially also interesting for the study of biopolymers.

\begin{acknowledgments}

The authors would like to acknowledge networking support by the EUTOPIA COST Action CA17139. This work has been partially supported by the ANR project ANR-19-CE45-0016. Vincent Dahirel and Jean-Marc Victor are warmly acknowledged for their constant support and insightful advice.

\end{acknowledgments}

\bibliography{biblioCompleteTim.bib}


\end{document}


\newcommand{\beginsupplement}{%
    \setcounter{table}{0}
    \renewcommand{\thetable}{S\arabic{table}}%
    \setcounter{figure}{0}
    \renewcommand{\thefigure}{S\arabic{figure}}%
 }

\beginsupplement
\onecolumngrid
\appendix

\renewcommand{\thefigure}{S\arabic{figure}}
\renewcommand{\theHfigure}{S\arabic{figure}}

\section{Supplementary information}

\subsection*{Coefficients \texorpdfstring{$a_1$}{a\_1}, \texorpdfstring{$a_2$}{a\_2} and \texorpdfstring{$\varepsilon_\theta$}{epsilon\_theta} for the three curves}

\begin{table}[h!]
\setlength{\tabcolsep}{20pt}
\renewcommand{\arraystretch}{2}
    \centering
    \begin{tabular}{|l|c c c|}
        \hline
        Data & ~~$a_1$~~ & ~~$a_2$~~ & ~~$\varepsilon_\theta$~~ \\
        \hline
        Inflection points  $\varepsilon^I(N)$ & $2.1  \pm 0.04$  &  $3.3 \pm 0.9$ & $0.2694\pm 0.0005$ \\
        Fluctuation maxima $\varepsilon^F(N)$ & $2.5  \pm 0.09$  &  $2.7 \pm 1.8$ & $0.270 \pm 0.002$ \\
        Random Walk $\varepsilon^{RW}(N)$     & $0.0  \pm 0.1$   &  $4.6 \pm 1.8$ & $0.269 \pm 0.003$ \\
        Vogel \textit{et al.}
        $\mathrm{d}C_v/\mathrm{d}t$
        [13]
        & $2.5$ & $8.0$  & $0.2690 \pm 0.0005$ \\
        \hline
    \end{tabular}
    \caption{In 
    Figure 2
    we have fitted inflection points (colored circles) and $\alpha_N(\varepsilon)$ fluctuation maxima (colored triangles) by the expression $\varepsilon_\theta(N) = {a_1}{{N^{-1/2}}} + {a_2}{(2N)^{-1
     }} + \varepsilon_\theta$ from Ref.~[13]
     (green and white lines, respectively). Here we list the corresponding fitting parameters $a_1$, $a_2$ and $\varepsilon_\theta$, together to those obtained by Vogel \textit{et al.} 
     in 
     Ref.~[13]
     by fitting a different order parameter.}
    \label{tab:Vogelfunction}
\end{table}

\subsection*{Reference values \texorpdfstring{$\alpha_C$}{alpha\_C}, \texorpdfstring{$\alpha_I$}{alpha\_I} and \texorpdfstring{$\alpha_G$}{alpha\_G} obtained for different \texorpdfstring{$M$}{M}}

\begin{table}[h]
\setlength{\tabcolsep}{20pt}
\renewcommand{\arraystretch}{2}
    \centering
    \begin{tabular}{|c|c|c|c|}
        \hline
        ~~$M$~~ & ~~$\alpha_C$\footnotemark [1]~~& ~~$\alpha_I$\footnotemark [1]~~ & ~~$\alpha_G$\footnotemark [1]~~ \\
        \hline
     3    & -1.81 & -1.09  & -0.36  \\
     4    & -2.03 & -1.10  & -0.19 \\
     5    & -2.11 & -1.10  & -0.11 \\
     6    & -2.15 & -1.10  & -0.07 \\
     7    & -2.16 & -1.10  & -0.05 \\
     8    & -2.19 & -1.11  & -0.03  \\
     10   & -2.19 &  -1.11  & -0.03 \\
     20   & -2.21 & -1.12  & -0.02 \\
        \hline
    \end{tabular}
    \footnotetext{Significant digits.}
    \caption{
    Numerical values of the loglog-slopes $\alpha_{N,M}(\varepsilon)$ obtained from the spectra in 
    Figure~3-a
    and displayed in 
    Figure~3-b
    (points), as a function of $M$.
    The parameter $\alpha_C$, for coils, is obtained at the limit $\varepsilon=0$; $\alpha_I$ is measured at the fitting sigmoid inflexion point; $\alpha_G$, for globules, is calculted as $\varepsilon \to \infty$ limit of the fitting sigmoid.
    }
    \label{tab:SuppalphaM}
\end{table}

\newpage
\subsection*{Critical \texorpdfstring{$\varepsilon^I(N)$}{epsilonI(N)} inflection point fit parameters upon decimation}

\begin{table}[h!]
\setlength{\tabcolsep}{20pt}
\renewcommand{\arraystretch}{2}
    \centering
    \begin{tabular}{|c| c | c | c|}
        \hline
        $M$ & ~~$a_1$~~ & ~~$a_2$~~ & ~~$\varepsilon_\theta$~~ \\
        \hline
         $3$  &$1.7 \pm 0.1$  & $2.7 \pm 1.8$  & $0.273 \pm 0.002$\\
         $5$  &$1.8 \pm 0.05$ & $5.0 \pm 0.9$  & $2.270 \pm 0.001$\\
         $8$  &$1.9 \pm 0.04$ & $4.4 \pm 0.7$  & $2.270 \pm 0.001$ \\ 
         $10$ &$2.0 \pm 0.05$ & $4.2 \pm 0.8$  & $0.270 \pm 0.001$ \\
        \hline
        $M=N$ & $2.1 \pm 0.04$  &  $3.3 \pm 0.9$  &   $0.2694 \pm 0.0005$ \\ 
        \hline
    \end{tabular}
    \caption{
    As shown in 
    Figure 3,
    decimated configurations retain the main information about the polymer state and allows us to reconstruct the phase diagram in the $(\varepsilon, N)$ phase space.
    Moreover, inflection points (colored circles in
    Figure 3
    can be used to define a critical line also in decimated conditions, and fitted by the function $\varepsilon_\theta(N) = {a_1}{{N^{-1/2}}} + {a_2}{(2N)^{-1}} + \varepsilon_\theta$.
    Here we give corresponding values of the three fitting parameters for $M=3, 5$, $8$, $10$ and, for comparison, for the non decimated case $M=N$.
    Note that the asymptotic value $\varepsilon_\theta$ is essentially independent on decimation, ensuring a correct definition of the phase transition in the thermodynamic limit.
    The $a_1$ and $a_2$ parameters, that gives the $N$ dependence of the transition in finite-size, are instead slightly modified by the decimation procedure, but converge toward the original values as soon as $M \gtrsim 10$.
    }
    \label{tab:Lines_with_M}
\end{table}

\newpage

\subsection*{Typical PSD at  the theta conditions \texorpdfstring{$\varepsilon = \varepsilon^{RW}_\theta(N)$ } {epsilon = epsilonRW(N)}}

\begin{figure}[h!]
    \centering
    \includegraphics[width=.8\textwidth]{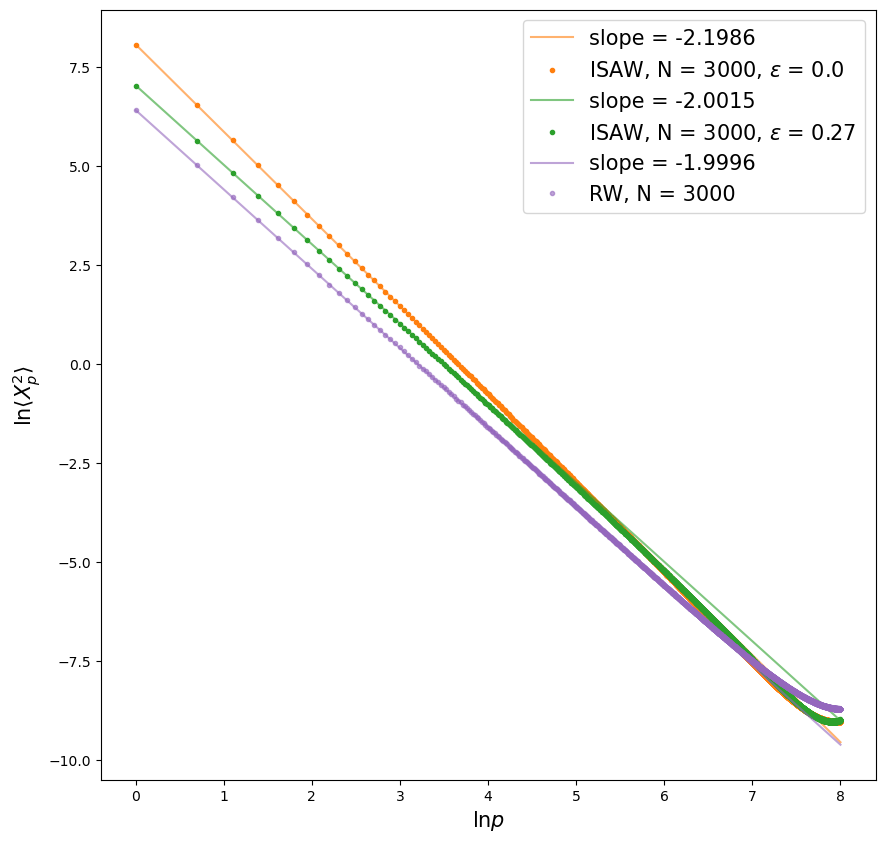}
   \caption{The plot shows a typical PSD obtained for a value of $\varepsilon$ matching the $\alpha = -2$ conditions, i.e for $\varepsilon = \varepsilon^{RW}_\theta(N)$ (green points). For comparison, a pure coil spectrum is shown (orange points) with its typical -2.2 slope (orange line) and a pure on-lattice RW (purple points) of slope -2 (purple line) are shown. The first modes slope $\alpha$ for $\varepsilon = \varepsilon^{RW}_\theta(N)$ is as expected equal to $-2$ (green line); Moreover, the plot shows how the same scaling law holds over a wide range of modes. This ensures that the polymer configurations can be described as random walks up to short scales (where a SAW behavior is recovered).}
    \label{fig:RW_PSD}
\end{figure}

\newpage

\subsection*{Effect of different levels of decimation on the phase transition}

\begin{figure}[h!]
    \centering
    \includegraphics[width=.8\textwidth]{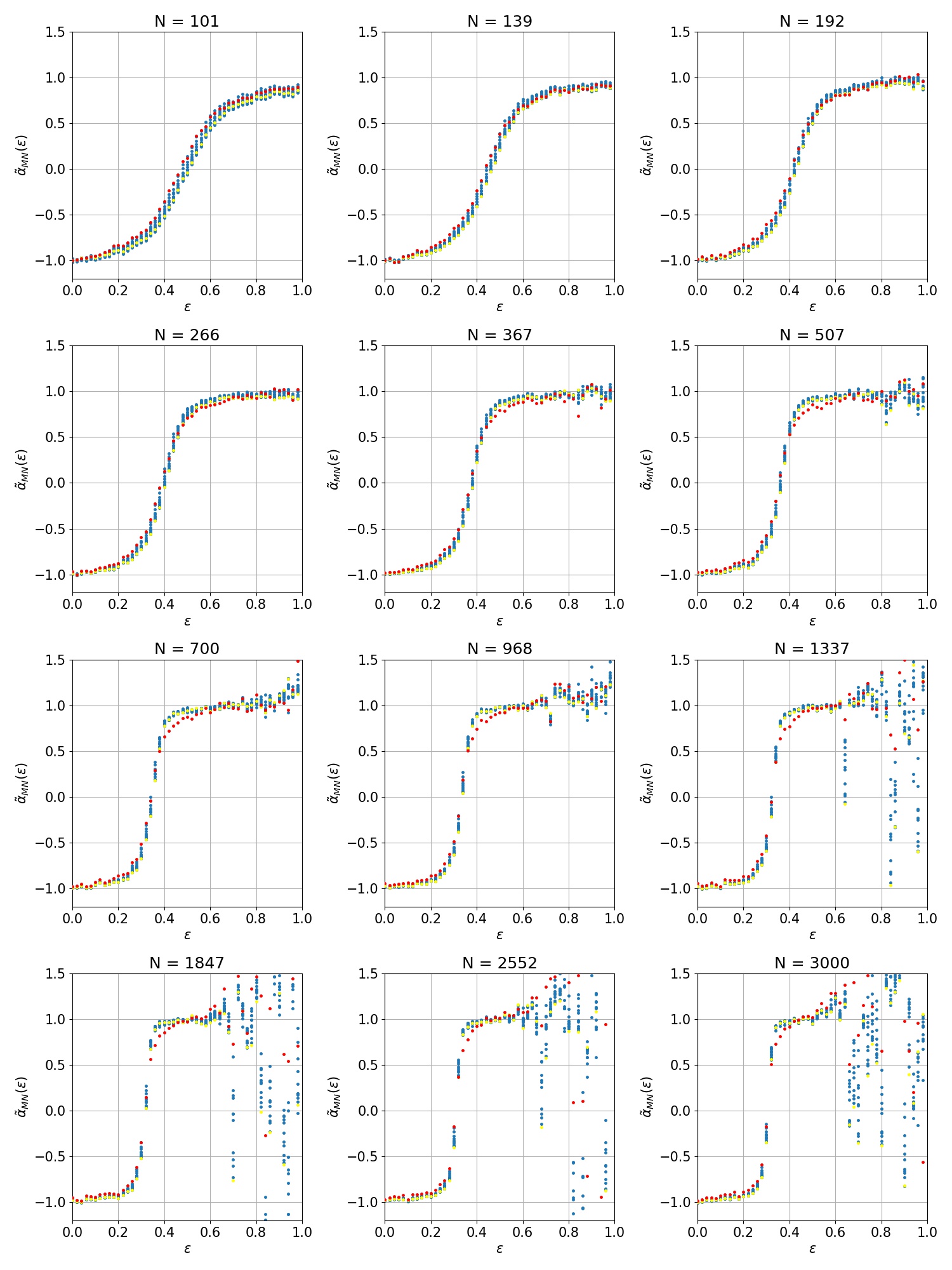}
    \caption{
    The different panels show renormalized loglog-slopes $\tilde{\alpha}_N(\varepsilon)$
    (Eq.~(6))
    for different value of $N$.
    For each $N$, we show results obtained by different levels of decimation $M$ (blue points), starting from 3 (red points) to $N$ (yellow points).
    Comparison of these different cases shows that, for given $N$, there is essentially no difference in the behavior of the order parameter ${\alpha}_N(\varepsilon)$, once rescaled with the maximum and minimum values, $\alpha_G(M)$ and $\alpha_C(M)$, respectively. Data sample includes here 16384 configurations per point.
    We recall that, as in
    Figure~2,
    samples from the sectors $(N, \varepsilon) \in [700;3000] \times [0.55;1.0]$ and $ [300;700] \times [0.7;1.0]$ didn't meet statistical relevance.
    Here we have kept the raw results from these sectors, which also gives a measure of the variability of the corresponding values.
    }
    \label{fig:sigmoids_several_M}
\end{figure}

\subsection*{Sigmoids and phase portrait from the $M=3$ decimated configurations}

\begin{figure}[h!]
    \centering
    \includegraphics [width=1\textwidth]{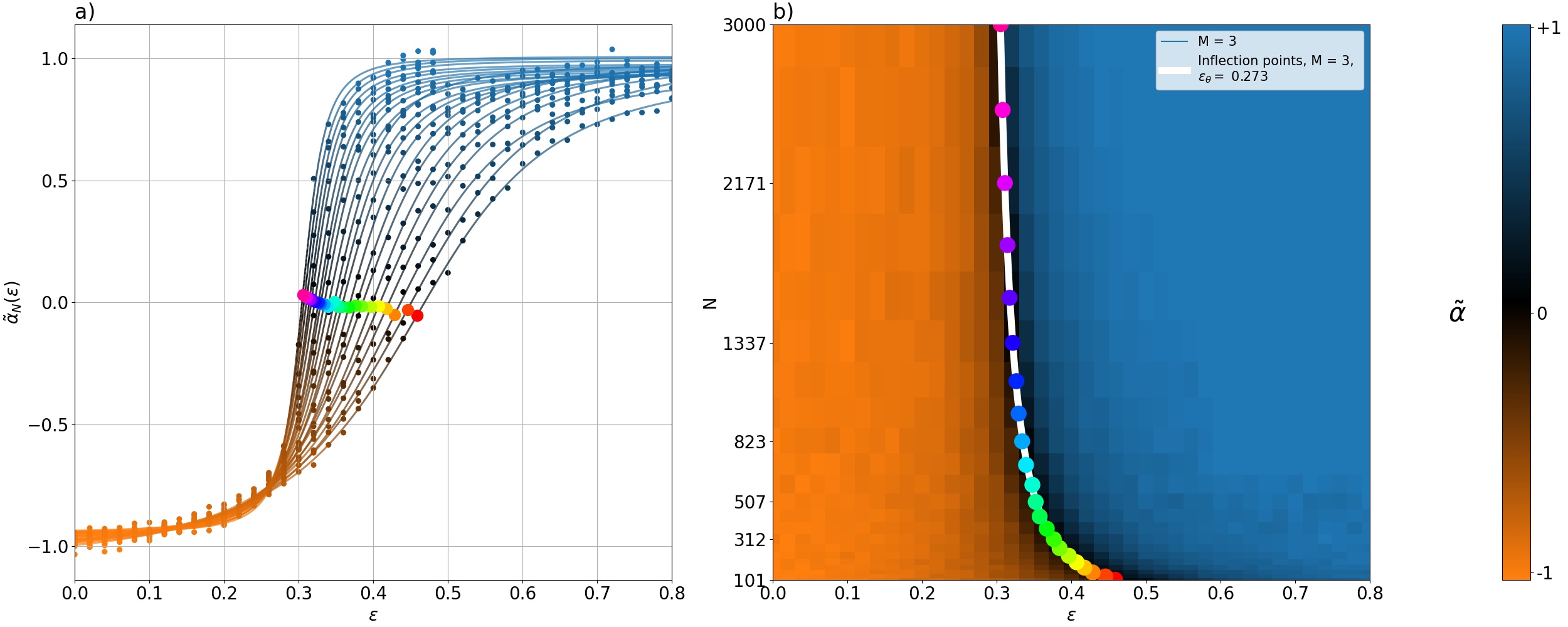}
    \caption{
    In Figure 3 of the main article are shown the Power Spectral Densities and critical values of $\alpha_{N,M}$ obtained by successive decimation of a $N= 431$ chain. Here we show, for the same system, 
    \textbf{(a)} the renormalized sigmoids $\tilde{\alpha}_{N,3}(\varepsilon)$ for $N$ from 101 to 3000 with decimation level $M=3$ ; and 
    \textbf{(b)} the phase portrait as reconstructed from the $M=3$ decimated signal. Colored dots are obtained from inflection points as in Figure 2. The white line is a fit yielding $\epsilon_\theta = 0.268$ (see Table S3). In both graphs we used a dataset of 16384 configurations.
    }
    \label{fig:decimation}
\end{figure}
\vspace{3cm}

\newpage

\newpage
\subsection*{Phase portraits after decimation, with different sample size.}

\begin{figure}[h!]
    \centering
    \includegraphics[width=1\textwidth]{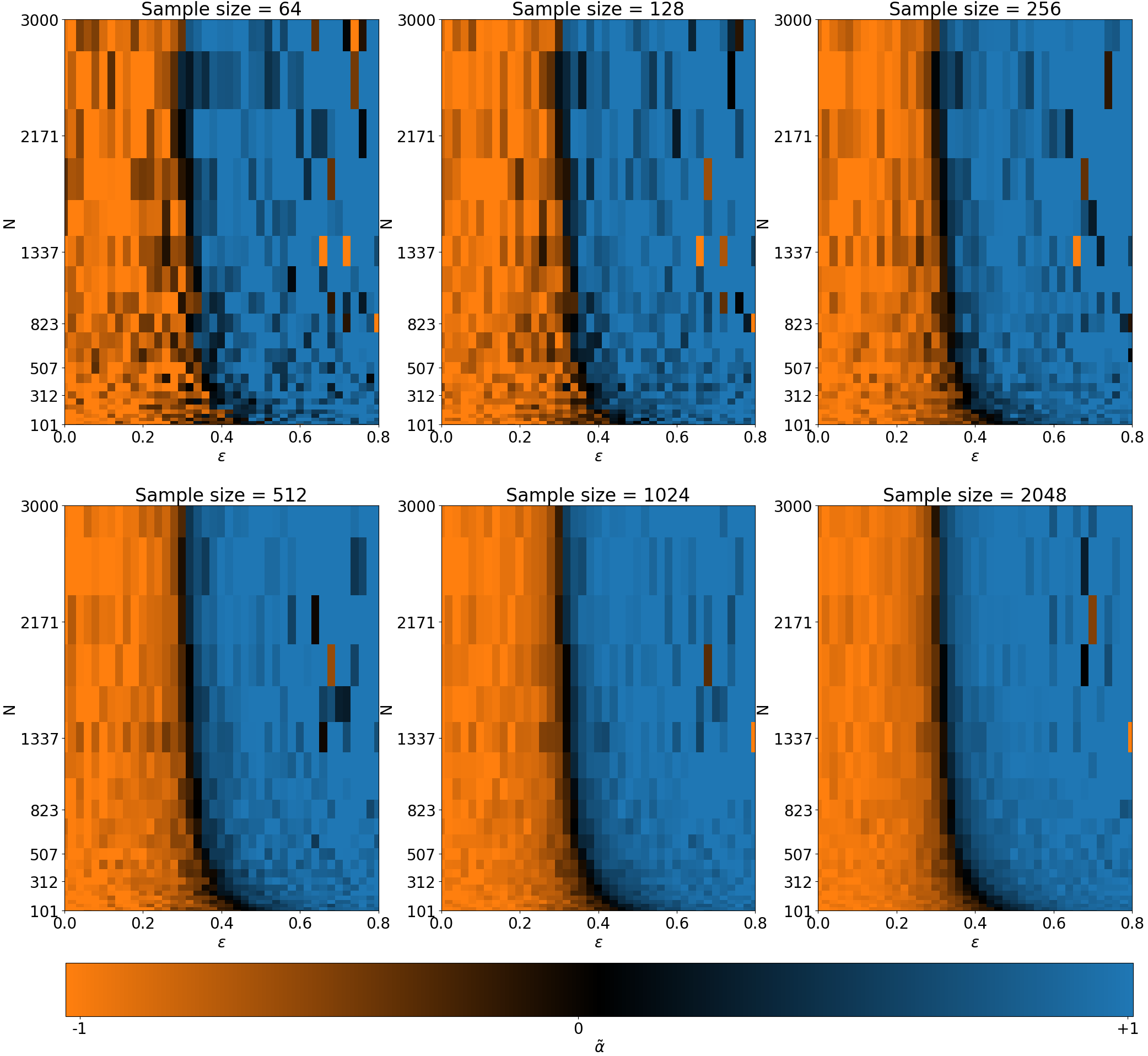}
    \caption{Example of phase portraits obtained by extreme decimation $M=3$ starting from a same set of conformations, for different \textit{sample} sizes, from $64$ to $2048$. We recall that, as in
    Figure 2,
    samples from the sectors $(N, \varepsilon) \in [700;3000] \times [0.55;1.0]$ and $ [300;700] \times [0.7;1.0]$ didn't meet statistical relevance. Here we have kept the raw results from these sectors, which also gives a measure of the variability of the corresponding values. Overall, we observe that if a sample of only 64 configurations is not sufficient to assign with certainty a system to the coil or globule states, it becomes rather good starting from 512.}
    \label{fig:Portraits_stats}
\end{figure}

\newpage

\subsection*{Simulation method details}

We simulated single ISAW on the cubic lattice and sampled conformations thanks to the Metropolis algorithm with reptation moves. We performed simulations for 22 values of $N$ ranging from 100 to 3000, and 100 values of $\varepsilon$ in the range  0.0--0.99. 
For each  $(N, \varepsilon)$, 32000 statistically independent conformations are recorded. Figure 1-a
shows typical conformations for the case $N=431$ and three different $\varepsilon$.
The three spatial components are DCT transformed, and their square amplitudes are summed to obtain their DCT spectra, as shown in Figure 1-c.
For any given $(N, \varepsilon)$, we average over the set of configurations to obtain the corresponding PSD.

Since the deep globular phase is dominated by low energy, entropically suppressed conformations, a statistical analysis in the high $N$, low $\varepsilon$ region requires sophisticated simulation methods that weren't undertaken in this study. 
Samples from the sectors $(N, \varepsilon) \in [700;3000] \times [0.55;1.0]$ and $ [300;700] \times [0.7;1.0]$ didn't meet statistical relevance and were omitted from the results.
